\documentclass[a4paper,12pt]{article}

\usepackage{graphicx}
\usepackage[left=2.5cm,right=2.5cm,top=2cm,bottom=2cm]{geometry}
\usepackage{subfig}  

\usepackage{xspace} 
\usepackage{amsbsy} 
\usepackage[version=3]{mhchem}

\newcommand{\celsius}{\ensuremath{\mathrm{^oC}}\xspace} 
\newcommand{\dg}{\ensuremath{\mathrm{^o}}\xspace}         
\newcommand{\gammaprime}{\ensuremath{\gamma^{\prime}}\xspace}
\newcommand{\gammachem}{\ce{AlAg2}}
\newcommand{\axis}[2][\alpha]{\ensuremath{\mathrm{[#2]_{#1}}}\xspace} 
\newcommand{\axes}[2][\alpha]{\ensuremath{\mathrm{\langle #2 \rangle_{#1}}}\xspace} 
\newcommand{\plane}[2][\alpha]{\ensuremath{\mathrm{(#2)_{#1}}}\xspace} 
\newcommand{\planes}[2][\alpha]{\ensuremath{\mathrm{\{#2\}_{#1}}}\xspace} 

\newcommand{\micron}{\ensuremath{\mathrm{\mu m}}\xspace}
\newcommand{\stdfig}[1]{\resizebox{7.25cm}{!}{\includegraphics{#1}}}
\newcommand{\medfig}[1]{\resizebox{08cm}{!}{\includegraphics{#1}}}

\newcommand{\burgers}[3]{\ensuremath{\boldsymbol{b}=\frac{#1}{#2}\langle #3 \rangle}\xspace} 
\newcommand{\burger}[3]{\ensuremath{\boldsymbol{b}=\frac{#1}{#2} [ #3 ]  }\xspace} 

\newcommand{\alagA}{\ensuremath{\mathrm{Al_{98.3}Ag_{1.7}}}\xspace}
\newcommand{\alagB}{\ensuremath{\mathrm{Al_{99.2}Ag_{0.8}}}\xspace}

\usepackage{fancyhdr}
\fancypagestyle{plain}{ %
\fancyhf{} 
\lhead{\textit{Precipitate assemblies in Al-Ag}}
\rhead{J. M. Rosalie, L. Bourgeois \& B. C. Muddle}
\lfoot{\textit{Pre-print}}
\rfoot{\thepage}
}

\begin{document}
\title{Precipitate assemblies formed on dislocation loops in aluminium-silver alloys}
\author{Julian~M.~Rosalie$^{a,b,c,\dagger}$
\and Laure~Bourgeois$^{b,c,d}$ \and
Barrington~C.~Muddle$^{b,c}$}
\date{}

\maketitle
\noindent
$^a$Microsctructure Design Group, Structural Metals Center, National Institute for Materials Science (NIMS), Japan.\\
$^b$ARC Centre of Excellence for Design in Light Metals, Australia.\\
$^c$Department of Materials Engineering, Monash University, 3800, Victoria, Australia.\\
$^d$Monash Centre for Electron Microscopy, Monash University, 3800, Victoria, Australia.
$^\dagger$Present address, work carried out at $\mathrm{^c}$.

\begin{abstract}
A detailed study of the precipitation of the \gammaprime (\gammachem) phase in undeformed aluminium-silver alloys has been conducted. 
Several previously unreported features were observed,  
including the formation of discrete three-dimensional assemblies comprised of 5-7 precipitates on faulted dislocation loops. 
The precipitates assemblies adopted a morphology approximating a tetrahedral  bipyramidal assembly.  
The bounding defect of the stacking fault was found to control both the nucleation of additional precipitates which was responsible for the formation of the assembly structure and also the growth characteristics of individual precipitates with these assemblies. 
\end{abstract}

\section*{Introduction}
\addcontentsline{toc}{section}{Introduction}

The \gammaprime (\gammachem) phase is a metastable 
 intermetallic precipitate formed in aluminium-silver alloys.
This alloy system has been extensively studied, 
principally on account of the structurally simple face-centred cubic to hexagonal close-packed 
(fcc\(\rightarrow\)hcp) 
phase transformation which produces the \gammaprime phase.   
As such it provides an important model system for several industrially significant aluminium alloy systems (e.g. Al-Cu-Mg) in which other plate-like precipitates serve as effective strengthening agents. 
An understanding of the solid-solid phase transformations required to form such plate-like precpitates is critical to the ongoing development and application of aluminium alloys, 
which are of particular importance within the transport and aerospace industries. 

The \gammaprime (\gammachem) phase forms as thin plates on the \planes[]{111} planes in aluminium-silver alloys aged at moderate temperatures \cite{voss:1999,frank:1961, nicholson:1961,passoja:1971}.
The \gammaprime unit cell is hexagonal, with lattice parameters 
\(a = 0.2858\)\,nm and \(c = 0.4607\)\,nm\cite{howe:1987,muddle:1994a}.
The \gammaprime phase adopts an orientation relationship of 
\(\plane{111}\parallel\plane[\gammaprime]{0001}; \axis{110}\parallel\axis[\gammaprime]{11\overline{2}0}\) to the aluminium matrix (\(\alpha\)) \cite{zakharova:1966}, 
with exceptionally good matching between the  \planes[]{111} matrix planes 
(\(2d_{111}=0.4687\)\,nm) 
and the \planes[]{0001} basal plane 
(\(d_{0001}=0.4607\)\,nm) 
of the precipitate.

The precipitation of the \gammaprime phase is particularly unusual in that it is accompanied by a very low 
\((-2.5\%)\) 
change in volume\cite{muddle:1994a}. 
This is in part due to the virtually identical atomic radii of silver and aluminium\cite{axon:1948} as well as to the structural similarities between the parent (face-centred cubic) and product (hexagonal close-packed) phases. 
The broad faces of the precipitate plates are coherent with the matrix, 
while the edges are semi-coherent with a misfit corresponding to single or multiple Shockley partial dislocations.
The structural change accompanying lengthening of the phase can be described by the glide of these bounding dislocations through the matrix, 
resulting in the extension of the hcp region\cite{muddle:1994a}.

The low volumetric strain for the precipitation of the \gammaprime phase, 
together with the coherent broad faces and low estimates for the interfacial energy
\cite{ramanujan:1992},
would suggest that precipitation should proceed readily. 
However, silver-rich Guinier-Preston zones are known to survive 120\,h of ageing at 160\celsius\cite{nicholson:1961}, 
indicating that significant solute remains in the matrix rather than partitioning to the \gammaprime phase.
Therefore it appears that the nucleation of the \gammaprime phase does not occur readily without the availability of heterogeneous nucleation sites.
Experimental results obtained to date\cite{frank:1961, nicholson:1961,passoja:1971} bear this out.

Nucleation of the \gammaprime (\gammachem) phase has
 been associated with a number of different nucleation sites, 
including grain boundaries \cite{clark:1967}, dislocations \cite{frank:1961, nicholson:1961, passoja:1971} and dislocation loops \cite{frank:1961}. 
The common features for nucleation on dislocations and dislocation loops appears to be the presence of a stacking fault at the dislocation site. 
It has been suggested that such a defect provides a suitable template region or nucleus for the hexagonal crystal structure of the precipitate phase
\cite{frank:1961, nicholson:1961}.
However, unambiguous evidence to support this has has been lacking.

Precipitation of the \gammaprime phase at vacancy condensation dislocation loops has been examined previously by Nicholson and Nutting\cite{nicholson:1961} and Frank \textit{et al.}\cite{frank:1961}. 
In the as-quenched state,
the alloy contained hexagonal dislocation loops which displayed stacking fault contrast.
It was concluded that these defects were Frank loops with Burgers vector \burgers{1}{3}{111},
which were identified as  the nucleation site of the  \gammaprime phase. 
The extinction distance for the stacking fault fringes of the loop was measured as  \(210\mathrm{\AA}\) for the \(\boldsymbol{g}=111\) reflection\cite{frank:1961}.    
This was taken as evidence that silver had segregated to the defect, as this value was considerably closer to the expected distance for silver \((250\mathrm{\AA})\) than that for aluminium \((646\mathrm{\AA})\).
 
The dislocation loops formed in aluminium-silver alloys were later characterised in greater detail\cite{westmacott:1971} and were indeed found to be Frank dislocation loops lying on the \planes[]{111} planes of the matrix. 
The loops were faceted along \planes[]{011} planes
and it was expected that the bounding Frank partial dislocation (\burgers{1}{3}{111}) would be dissociated to form Shockley (\burgers{1}{6}{112}) and stair-rod
(\burgers{1}{6}{110}) partial dislocations. 
The dissociation reaction replaces a dislocation of large
\( \left( a \frac{\sqrt{3}}{3}\right) \)
 Burger's vector with two dislocations with shorter Burgers vectors, 
leading to a reduction in the overall elastic strain energy.

Later work has focused largely on the nucleation and growth of the \gammaprime phase on 
linear dislocations and dislocation helices (see, for example \cite{passoja:1971}). 
Dislocation helices are reported to be the predominant nucleation site in Al-16wt\%Ag, 
with dislocation loops playing a lesser role\cite{nicholson:1961}.
The nucleation of \gammaprime on dissociated perfect dislocations has been described \cite{voss:1999}, 
but the samples used were subjected to heavy cyclic deformation and dislocation loops were not observed.

The confusion over the bounding dislocation of the dislocation loop in Al-Ag alloys 
and the emphasis on the nucleation of the \gammaprime phase in deformed samples has meant that precipitation on dislocation loops in Al-Ag alloys has not received attention in the literature. 
Therefore, there is a need for a re-examination of the precipitation of the \gammaprime phase in the absence of external deformation, 
where dislocation loops should be the primary heterogeneous site available to assist in the nucleation of the \gammaprime phase. 
Hence the present investigation, 
in particular a detailed examination of the microstructure of undeformed binary Al-1.68at.\%Ag 
and Al-0.84at\%Ag alloys, 
focusing on the formation of a complex three-dimensional assembly structure which has not been previously reported.

\section*{Experimental details}
\addcontentsline{toc}{section}{Experimental details}

The binary aluminium silver alloys used in this work were cast from high-purity aluminium (Cerac alloys, 99.99\% purity) 
and silver (AMAC alloys, 99.9+\%).
Two alloy compositions were prepared, containing 1.68at.\%Ag and 0.84at.\%Ag, respectively, 
in order to investigate the effect of varying amounts of silver. 
These alloys are hereafter designated \alagA and \alagB. 
The pure elements were melted in air at 700\celsius in a graphite crucible, 
then poured into graphite-coated steel molds. 
The cast ingots were homogenised at 525\celsius for 7\,days.
Ingots were later hot-rolled to 2\,mm (for hardness testing), 
or 0.5\,mm for TEM analysis.
Alloy compositions were confirmed by inductively coupled plasma atomic emission spectrometry (ICP-AES).

Solution treatments were carried out for 0.5\,h at 525\celsius in a salt pot.
Specimens were quenched into water at ambient temperature and then aged at 200\celsius using an oil bath for up to 72\,h.

Vickers hardness testing was carried out using loads of  2.5\,kg, 5\,kg or 10\,kg, depending on the hardness of the material. 
The consistency of results obtained at each load was compared, 
and the difference was roughly \(\pm1\) Vickers hardness number (VHN).

Foils for TEM analysis were jet electro-polished 
using a nitric acid/methanol solution 
(33\% \ce{HNO3}/ 67\%\ce{CH3OH} by volume) 
at temperatures of \(-20\celsius\). 
The voltage applied was -13\,V, 
and currents averaged 200\,mA. 

Conventional transmission electron microscopy investigations were carried out using a Philips CM20 transmission electron microscope operating at 200\,kV. 
High-resolution TEM studied were conducted using a JEOL~2011 instrument, 
also operating at 200\,kV, with a point resolution of 0.23\,nm.

\section*{Results}
\addcontentsline{toc}{section}{Results}

\begin{figure}[hbtp]
\begin{center}
\medfig{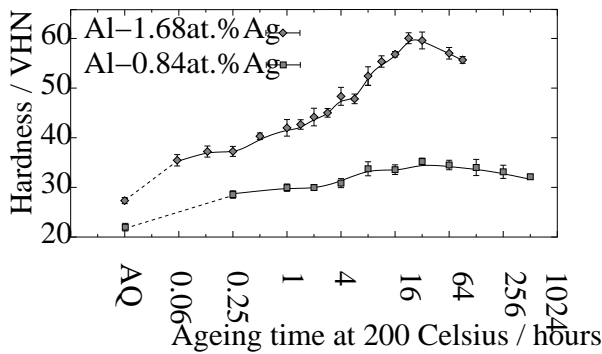}
\caption[Hardness curves for Al-Ag and Al-Ag-Cu alloys.]
{Hardness curves (VHN) for Al-Ag alloys  aged at 200\celsius.  
The curves are intended as a guide for the eye only. 
``AQ'' indicates the hardness in the as-quenched condition. 
\label{fig-hardness-curve1}}
\end{center}
\end{figure}

\paragraph{Hardness measurements}
Hardness curves obtained from Vickers hardness tests on the binary 
alloys are shown in Figure~\ref{fig-hardness-curve1}. 
The hardening response of the binary Al-Ag alloys was poor, 
with maximal hardness values of \(\mathrm{60\pm1\,VHN}\)  and 
\(\mathrm{35\pm1\,VHN}\) for \alagA and \alagB, respectively. 
Both alloys reached maximum hardness after 32\,h of ageing at 200\celsius. 

\subsection*{Transmission electron microscopy}

\paragraph{Quenched-in defect structure \label{para-alagA-aq}}

\begin{figure}
\begin{center}
\subfloat[\(\boldsymbol{g}=\mathrm{[0\overline{2}2]}\)]{\stdfig{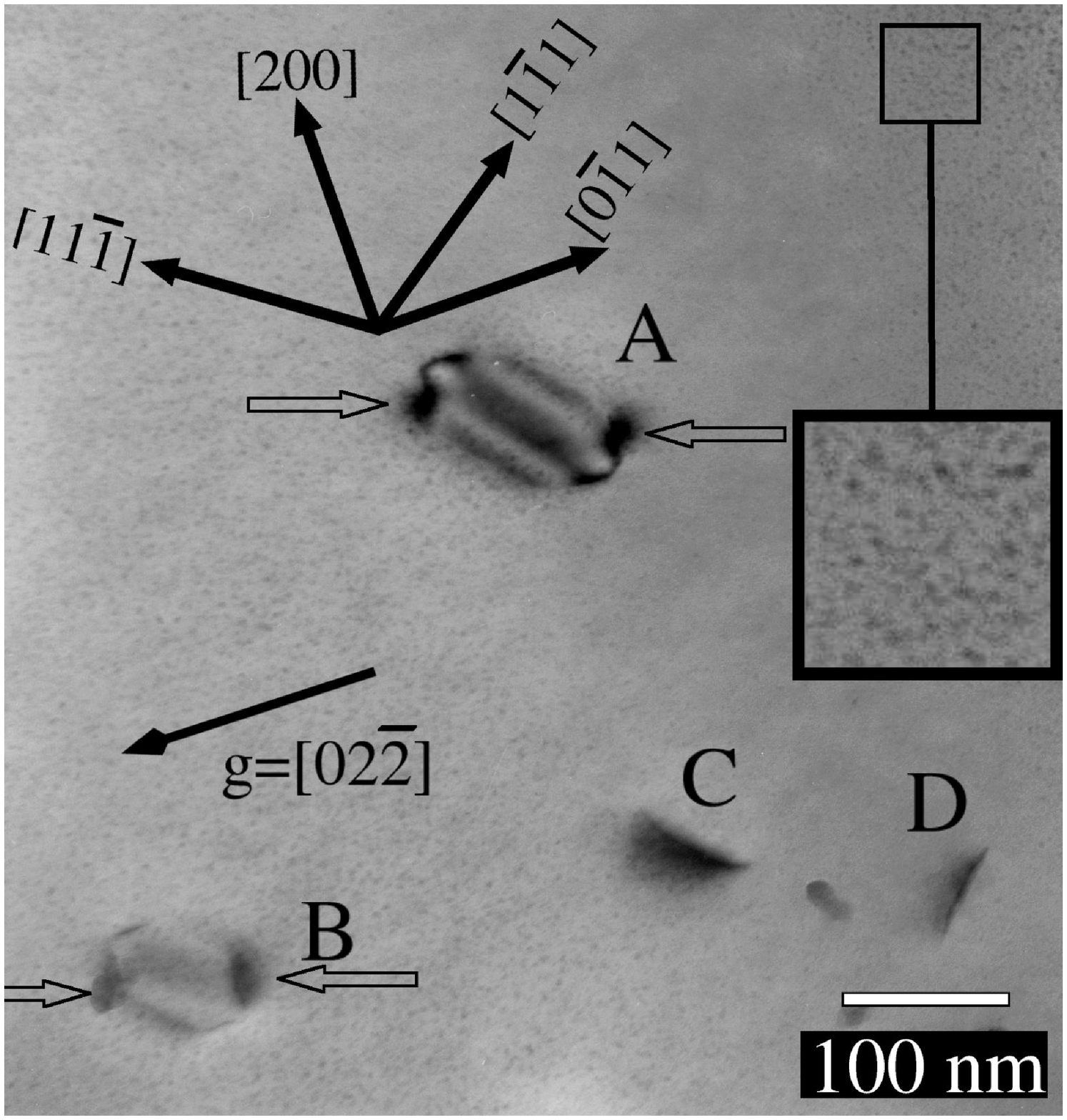}}
\hfill
\subfloat[\(\boldsymbol{g}=\mathrm{[02\overline{2}]}\)]{\stdfig{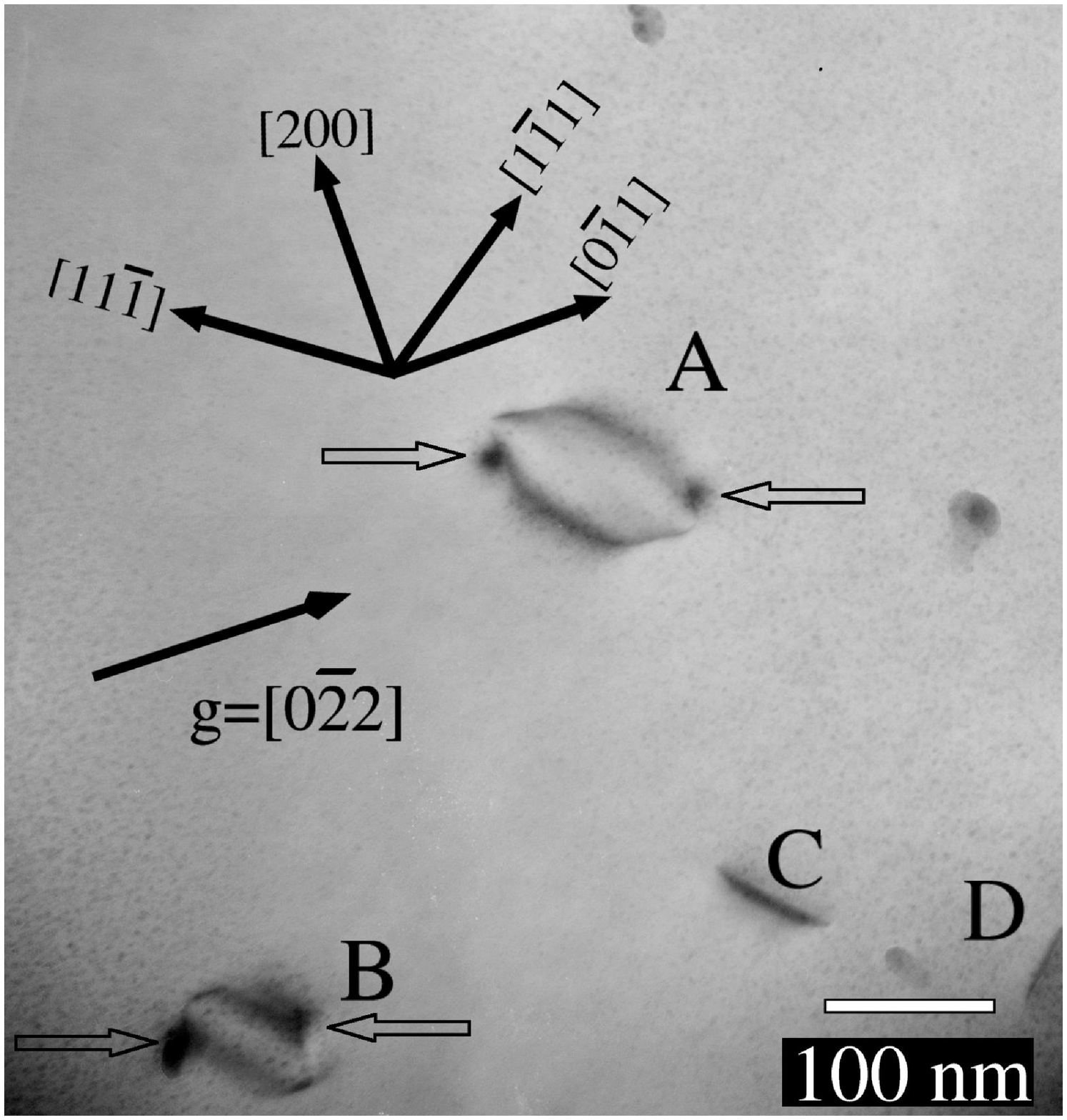}}
\caption[Electron micrographs showing dislocation loops in  the \alagA alloy after ageing for 0.083\,h at 200\celsius. ]
{Electron micrographs showing dislocation loops in  the \alagA alloy after ageing for 0.083\,h at 200\celsius. 
For loops at points \(A\) and \(B\) the dislocation line contrast switches from inside (in (a)) to outside (in (b)) for \(\mathrm{\mathbf{g} =\pm [0\overline{2}2]}\), however there is no reversal of the dislocation line strain contrast 
(indicated with hollow arrows). 
The latter feature is characteristic of a dissociated Frank loop \cite{clarebrough:1969,clarebrough:1969a}.
The presence of the fine-scale silver-rich spheroidal GP zones is shown in the enlarged inset in (a). 
\label{fig-5min-burgers}}
\end{center}
\end{figure}

The as-quenched defect structure was examined using two-beam diffraction contrast TEM.
Foils examined in this state contained a homogeneous distribution of dislocation loops and  spheroidal silver-rich Guinier-Preston (GP) zones.
The dislocation loops in \alagA had a diameter of approximately 50\,nm, while the GP zones had a diameter of approximately 1\,nm.
Brief ageing (0.083\,h at 200\celsius)  resulted in an increase in loop diameter, 
as shown in Figure~\ref{fig-5min-burgers}, 
assisting in the TEM analysis.
It was then possible to determine that hexagonal dislocation loops lay on the \planes[]{111} planes of the matrix, 
with facets parallel to \planes{110} planes and displayed stacking fault contrast for \(\boldsymbol{g}=111\) type reflections. 

The strain contrast along the dislocation line segments (indicated by the hollow arrows)  in a \(\mathbf{g}=\axes[]{02\overline{2}}\) condition (as shown in Figure~\ref{fig-5min-burgers}a), remained unchanged for \(\mathbf{g}=\axes[]{0\overline{2}2}\) (as shown in \ref{fig-5min-burgers}b).
The absence of a contrast reversal for reflections of type \(\mathrm{\mathbf{g} =\pm [0\overline{2}2]}\) 
has been shown  to indicate the dissociation of the Frank partial dislocation to form Shockley and stair-rod partial dislocations \cite{clarebrough:1969}. 
Note that the position of the contrast feature remains unchanged, 
even though the contrast within the loop switches from inside contrast (in Figure~\ref{fig-5min-burgers}a) to outside contrast (in Figure~\ref{fig-5min-burgers}b). 
The dislocation loops in \alagB alloys appeared to show similar features, 
however these were smaller in diameter and could not be fully characterised. 
The silver-rich GP zones can be seen in the enlarged inset in Figure~\ref{fig-5min-burgers}a.

\paragraph{Precipitate assembly microstructure \label{para-alagA-prec}}

\begin{figure}[h!]
\begin{center}
\subfloat[\label{precB}]
{\stdfig{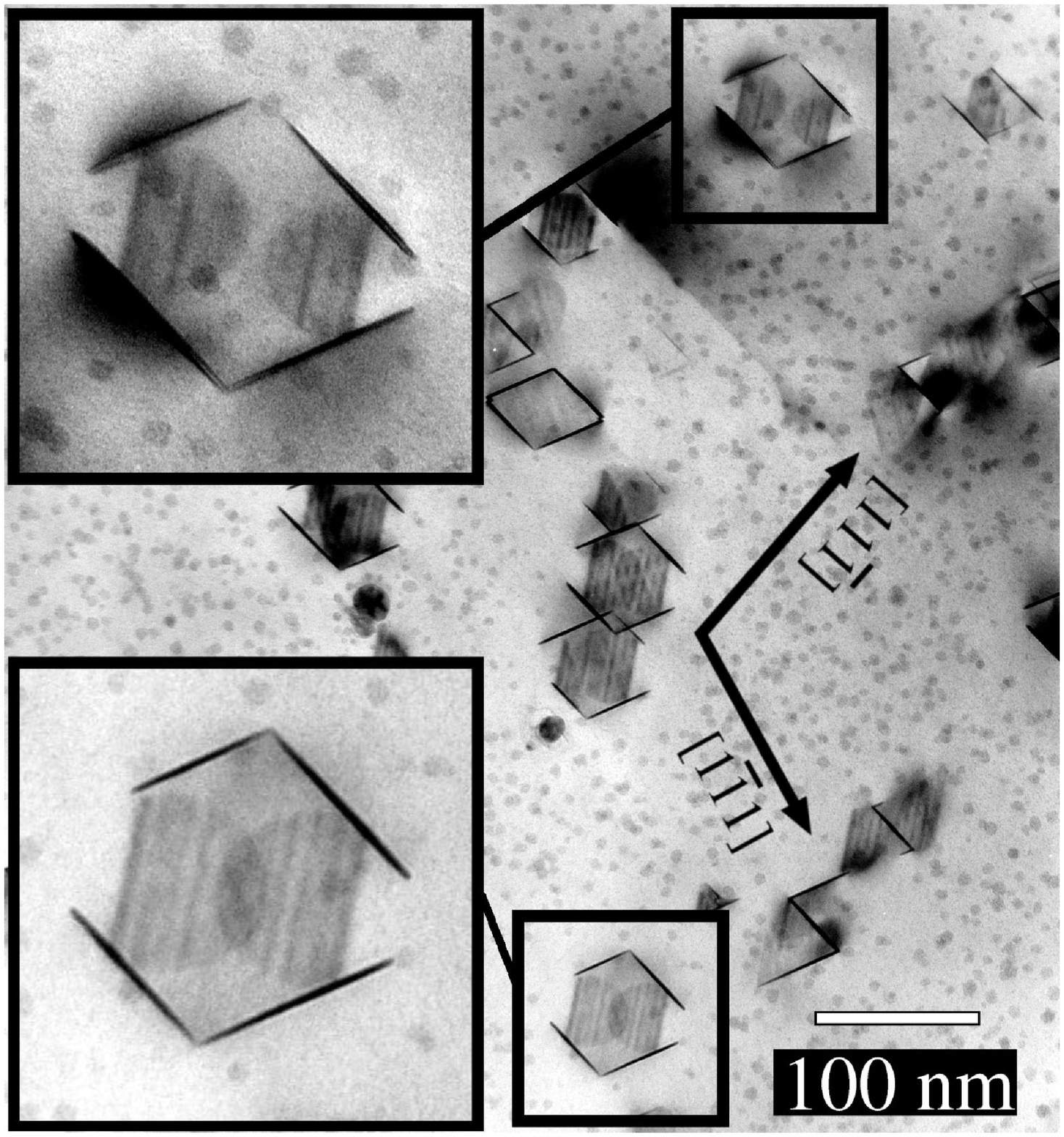}}
\hfill
\subfloat[\label{precC} ]
{\stdfig{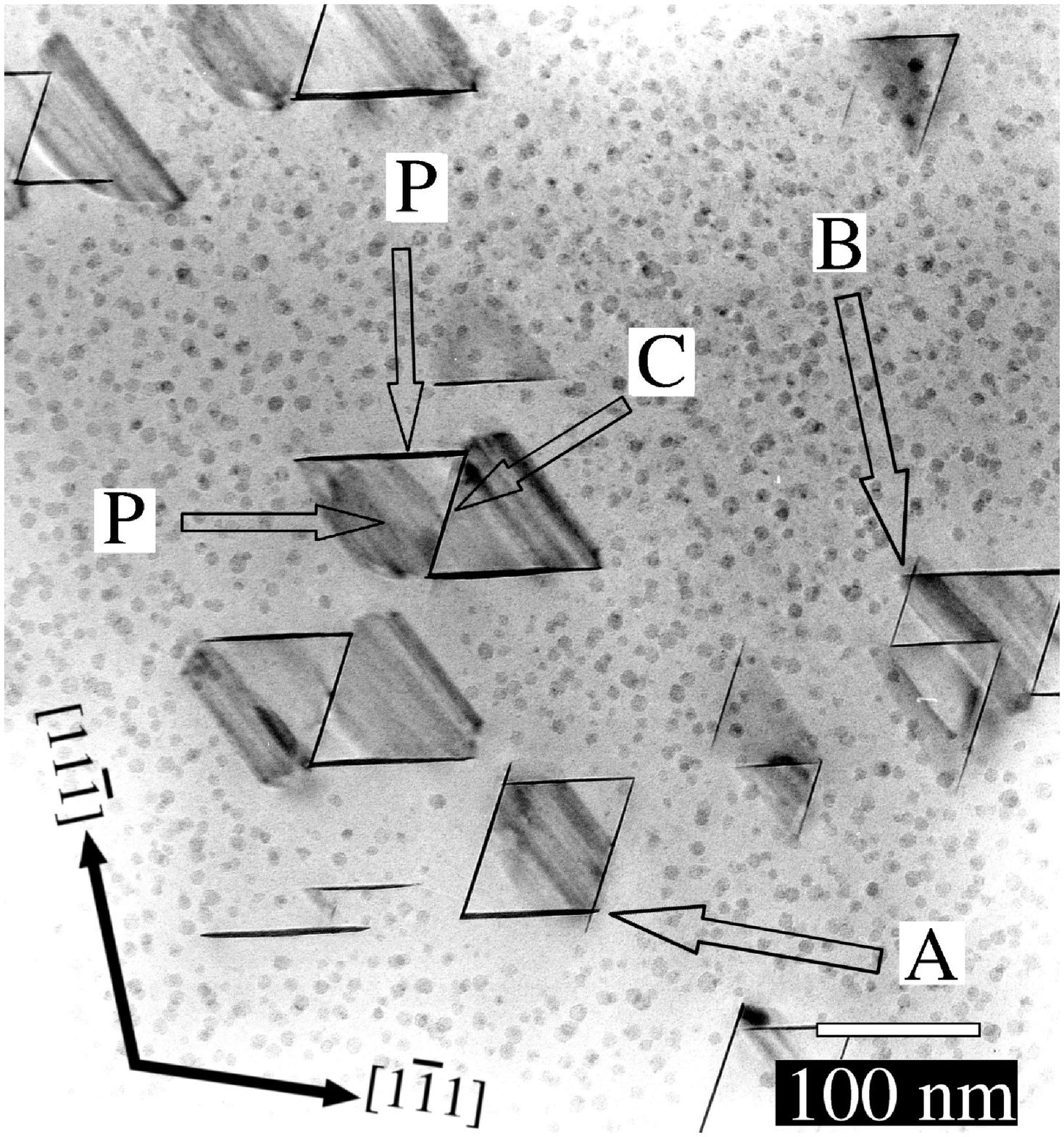}}

\caption{Micrographs of 
(a) the \alagB alloy after ageing for 2\,h 
 at 200\celsius, and
(b) the \alagA alloy after ageing for  
8\,h,
viewed parallel to \axis{011} zone axis.
The arrows indicate central (\(C\)) and peripheral (\(P\)) precipitates in the
assembly structure. 
Tilting experiments showed that the apparent overlap of precipitates at points \(A\) and \(B\) was an effect of projection. 
} 
\label{prec}
\end{center}
\end{figure}

Precipitation of \gammaprime was first observed after 0.5\,h of ageing at 200\celsius.
In this ageing condition, 
the microstructure contained spheroidal GP zones with diameters of approximately 1\,nm  
and thin \gammaprime plates on the \planes{111} planes. 
(See Figure~\ref{prec}). 
Two variants of the precipitate were perpendicular to the beam in this orientation, 
with a further two variants inclined to the beam. 
GP zones were absent from the volume close to the precipitates. 

Following further ageing, the precipitate was present primarily in small assembly-like aggregates, as shown in Figure~\ref{precB}, 
where the \alagB alloy has been aged for 2\,h at 200\celsius. 
Similar results were found for \alagA. 
The hexagonal central precipitate was surrounded by six \gammaprime precipitates (4 perpendicular to the electron beam and 2 inclined to the beam), occupying each of the \axes{011} edges of the central precipitate.
 
The peripheral plates continued to grow upon further ageing to sizes comparable to, 
or larger than, 
the central precipitates, 
as shown in Figure~\ref{precC}, 
which displays the microstructure observed after 8\,h of ageing in the \alagA alloy. 
The central plates, however,  
do not appear to have increased in diameter.
Tilting experiments showed that the apparent hard impingement between the peripheral precipitates at points \(A\) and \(B\) in Figure~\ref{precC} was an effect of viewing in projection.
Silver-rich GP zones were still observed after ageing up to 72\,hours.

\begin{figure}[h!]
\begin{center}
\stdfig{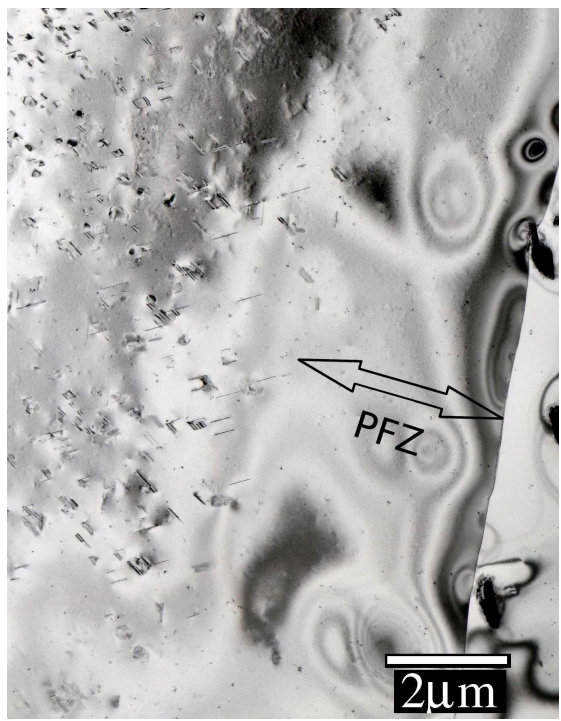}
\caption{A micrograph showing the grain boundary precipitate free zone 
(PFZ) in 
the \alagB alloy after ageing for  
2\,h.
} 
\label{pfz}
\end{center}
\end{figure}

The precipitate assemblies were widely distributed through the matrix, 
however,
their formation appeared to be sensitive to the presence of other defects. 
The assemblies were not formed within \(\sim2-3\)\,\micron of the grain boundary 
(See Figure~\ref{pfz}), for example, 
although GP zones were observed 50\,nm from the boundaries. 
The width of the grain boundary precipitate free zone (PFZ) differed between samples, possibly reflecting variations in the quenching rate. 
In addition to the presence of the aggregated precipitates, 
large \gammaprime precipitates were occasionally observed in association with either linear or helical dislocations.
Where this occurred, the surrounding volume of matrix showed a reduced density of the \gammaprime precipitate assemblies. 

\begin{figure}
\subfloat[\label{hrtem-1a}]{\stdfig{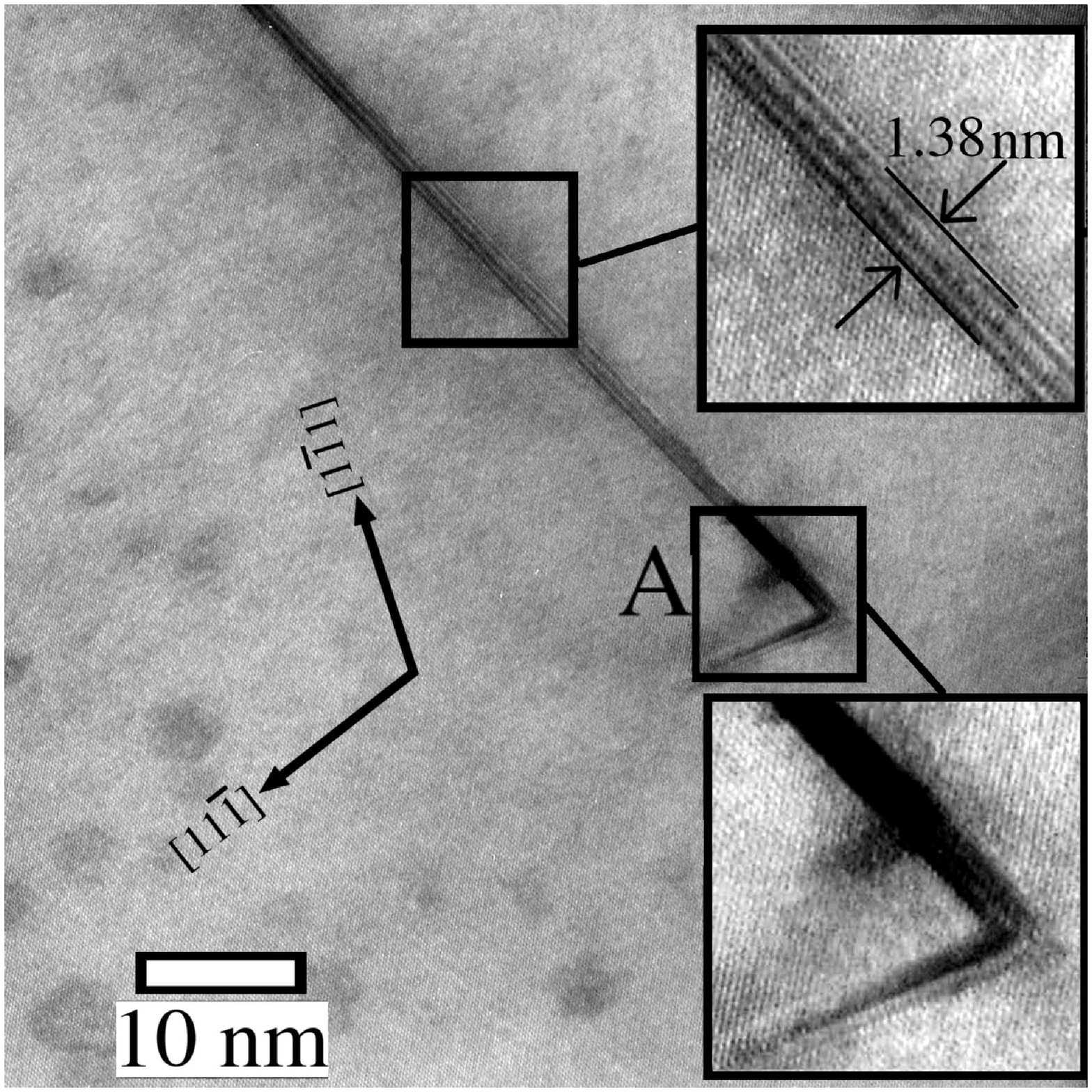}}
\hfill
\subfloat[\label{hrtem-1b}]{\stdfig{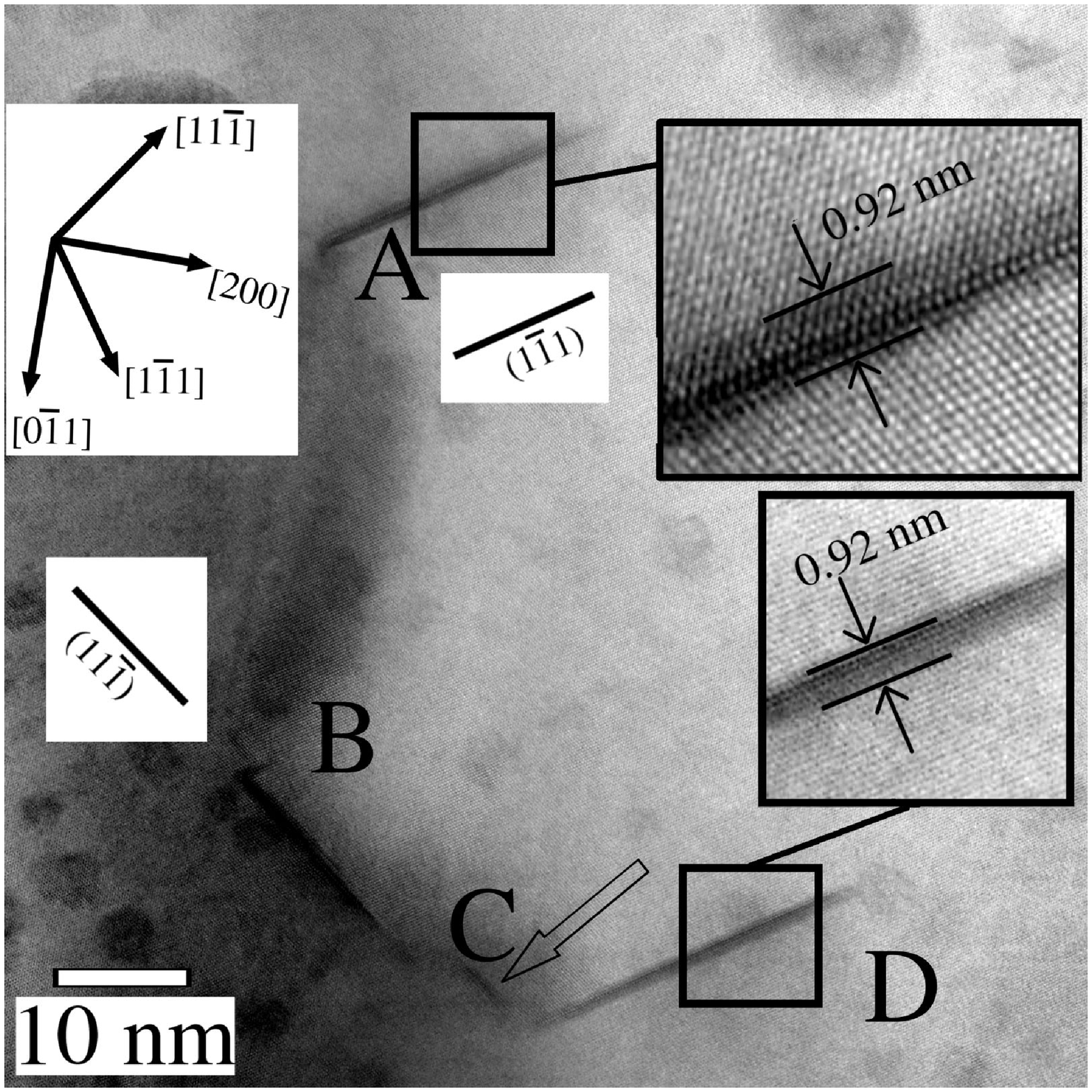}}
\caption{\label{hrtem-1}
High resolution TEM micrographs of precipitate assemblies in \alagA viewed along \axis{011}.
(a) Shows the central precipitate edge-on, 
with a peripheral precipitate present.
The region surrounding the precipitates is depleted in Ag-rich GP zones. 
An additional, smaller peripheral precipitate is present at \(A\) and
is shown in the enlarged inset. 
The central precipitate had a thickness of 1.38\,nm, 
while the peripheral precipitate close to \(A\) had a thickness of 0.92\,nm.
(b) Shows the peripheral plates of another assembly. 
The central plate is inclined to the beam and in poor contrast.
Peripheral precipitates are present at points \(A\) to \(D\).
Note that there are two peripheral plates (\(B\&C\)) on the 
\plane[]{11\overline{1}} plane, apparently occupying the same edge of the central 
precipitate.
The thickness of plates at \(A\) and \(D\) was 0.92\,nm.}
\end{figure}

\textbf{High-resolution transmission electron microscopy}
High resolution transmission electron microscopy (HRTEM) was used to examine the \gammaprime phase assemblies after brief ageing. 
The electron micrograph given in Figure~\ref{hrtem-1a} shows the early stages of the formation of a precipitate assembly in a foil of \alagA aged for 0.5\,h at 200\celsius. 
The micrograph shows the central precipitate 
with a peripheral plate (with diameter\(<10\)\,nm) located at its edge. 
The thicknesses of the central and peripheral plate were 1.38\,nm and 0.92\,nm,  corresponding to 3 and 2 unit cell heights of the \gammaprime unit cell, respectively.

Figure~\ref{hrtem-1b} shows a number of peripheral plates around the edges of a central plate. 
The latter is inclined to the beam direction and in poor contrast. 
Peripheral plates (with thickness of 0.92\,nm) can be seen at points \(A-D\).
Note that the precipitates at \(B\) and  \(C\) both lie on the \plane{11\overline{1}} plane, and apparently occupy the same edge of the central precipitate, 
but are slightly displaced from one another in the \axis{11\overline{1}} direction.

\section*{Discussion}
\addcontentsline{toc}{section}{Discussion}

\paragraph{The nucleation site of the \gammaprime precipitates} 

The structured assemblies of \gammaprime precipitates studied in this work were the primary form of intra-granular precipitation in the \alagA and \alagB alloy and appeared to form on hexagonal Frank dislocation loops. 
Pre-ageing deformation was deliberately avoided 
and while dislocation loops and dislocation helices were observed occasionally,
they were relatively uncommon.
Each dislocation loop can therefore act as an isolated nucleation site, 
without the additional complications of short-circuit diffusion of solute and/or vacancies along grain boundaries or dislocation helices.

The precipitate assemblies were not observed in the vicinity of vacancy sinks, 
including dislocation helices, 
dislocations and grain boundaries (as shown in Figure~\ref{pfz}). 
Each of these sites is expected to reduce the local vacancy population and thus inhibit the formation of dislocation loops in the surrounding volume. 

\begin{figure}
\begin{center}
\subfloat[]{\stdfig{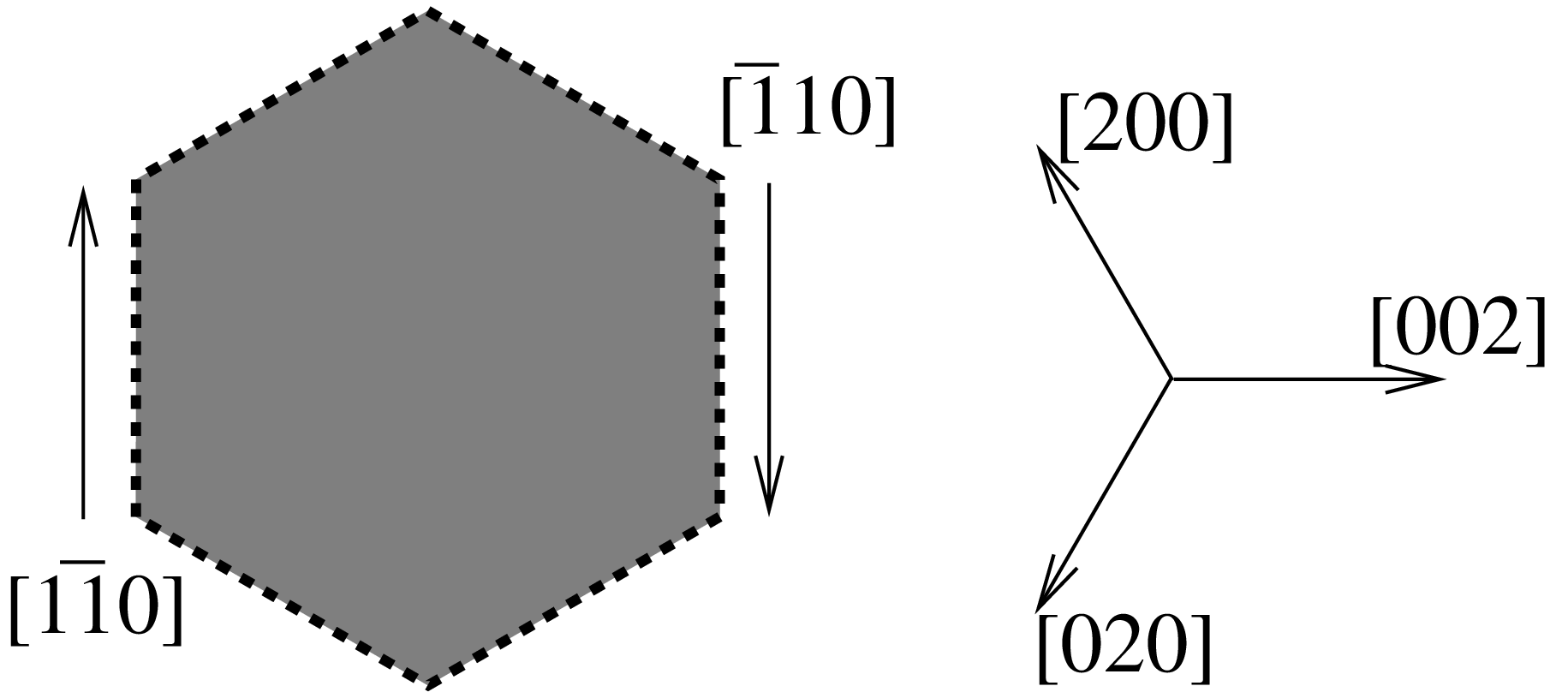}}
\hfill
\subfloat[\label{fig-frank-disoc-b}]{\stdfig{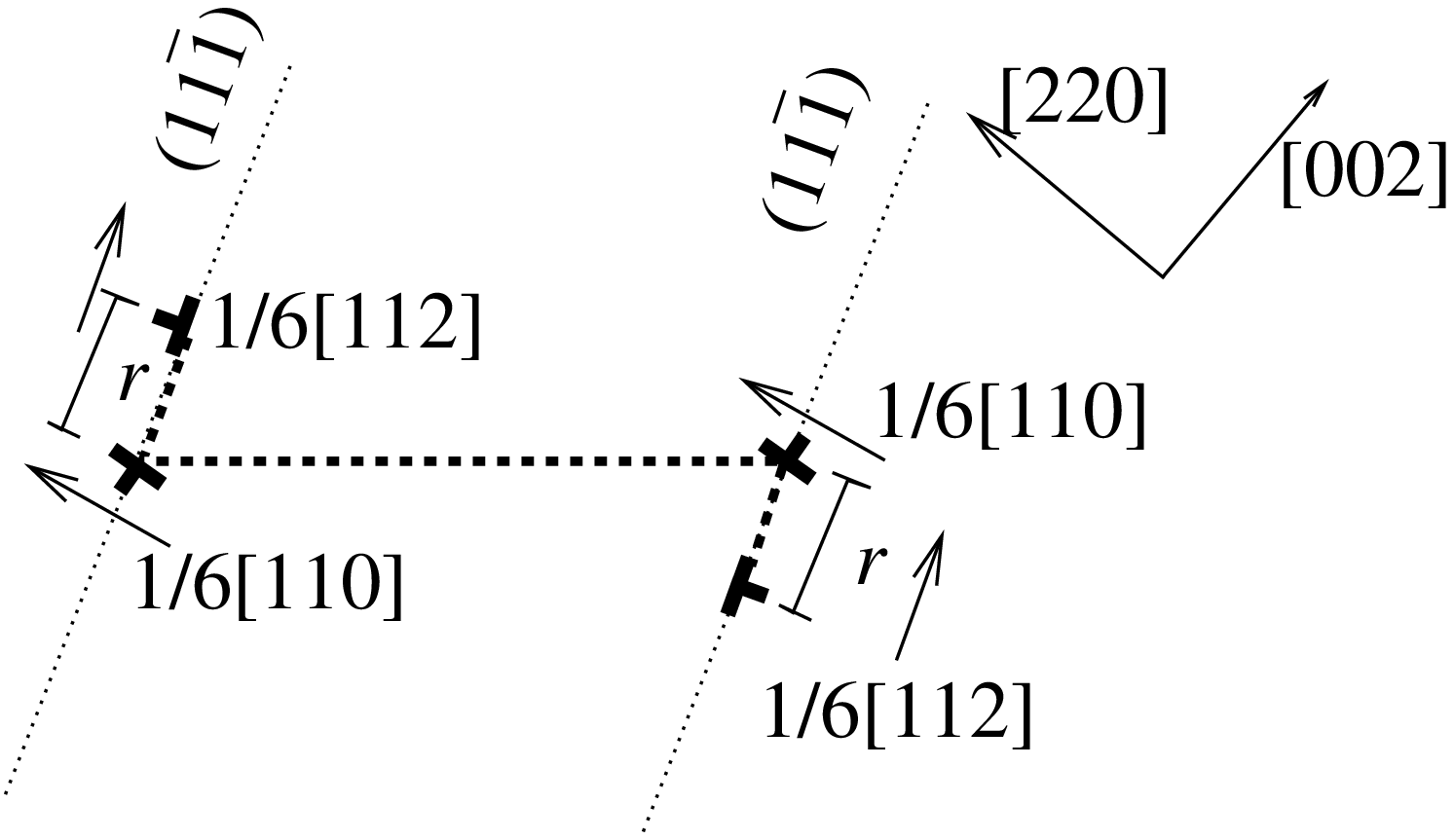}}

\caption[The dissociation of one face of a Frank loop.]
{The dissociation of a Frank loop. 
The hexagonal shape of the loop is shown in (a) where the loop is viewed along the \axis[]{111} direction  
normal to the \plane[]{111} plane on which the loop lies. 
The faulted region is indicated by the hashed region. 
(b) illustrates the dissociation of this dislocation into Shockley (\burger{a}{6}{112})
and stair-rod (\burger{a}{6}{110}) partial dislocations.
The 
\( (11\overline{1}) \) 
glide planes of the Shockley partial dislocations and the separation, \(r\), 
between the partial dislocations are also indicated. 
Note that in addition to the original faulted region, 
the passage of the Shockley partial dislocation generates a stacking fault. 
This process occurs for each \planes[]{110} facet in the habit plane of the loop, 
forming six Shockley partial dislocations which are located alternately above and below the plane of the loop. 
\label{fig-frank-disoc}}
\end{center}
\end{figure}

TEM observations indicated that the Frank loops were dissociated. 
Two-beam diffraction TEM indicated that the bounding Frank (\burger{a}{3}{111}) dislocations had dissociated into stair-rod and Shockley partial dislocations (See Figure~\ref{fig-5min-burgers}). 
Each of the Shockley partial dislocations lies on a different \planes[]{111} plane to the habit plane of the loop and is able to glide on that plane, 
whereas the stair-rod dislocation is sessile. 
The glide of the Shockley partial dislocations results in the formation of an additional region of stacking fault, 
as illustrated in Figure~\ref{fig-frank-disoc}. 
The \gammaprime phase forms on the faulted (hcp) surface provided by the Frank loop, both in the plane of the loop, and between the Shockley and stair-rod partial dislocations.
This process is repeated at each edge of the hexagonal Frank loop to yield the assembly structures observed in these alloys.

\paragraph{Precipitation assembly structure}
The predominant form of \gammaprime phase precipitation was as three-dimensional assembly structures
(Figure~\ref{prec}). 
 These were comprised of a single, hexagonal central precipitate with up to six peripheral precipitates, each sharing a common edge with the central plate.
The HRTEM results indicate that these structures are formed very early in the ageing process and were observed for precipitates of diameters \(<10\)\,nm and thicknesses corresponding to two unit cell heights of \gammaprime.

\begin{figure}[hptb]
\begin{center}
\stdfig{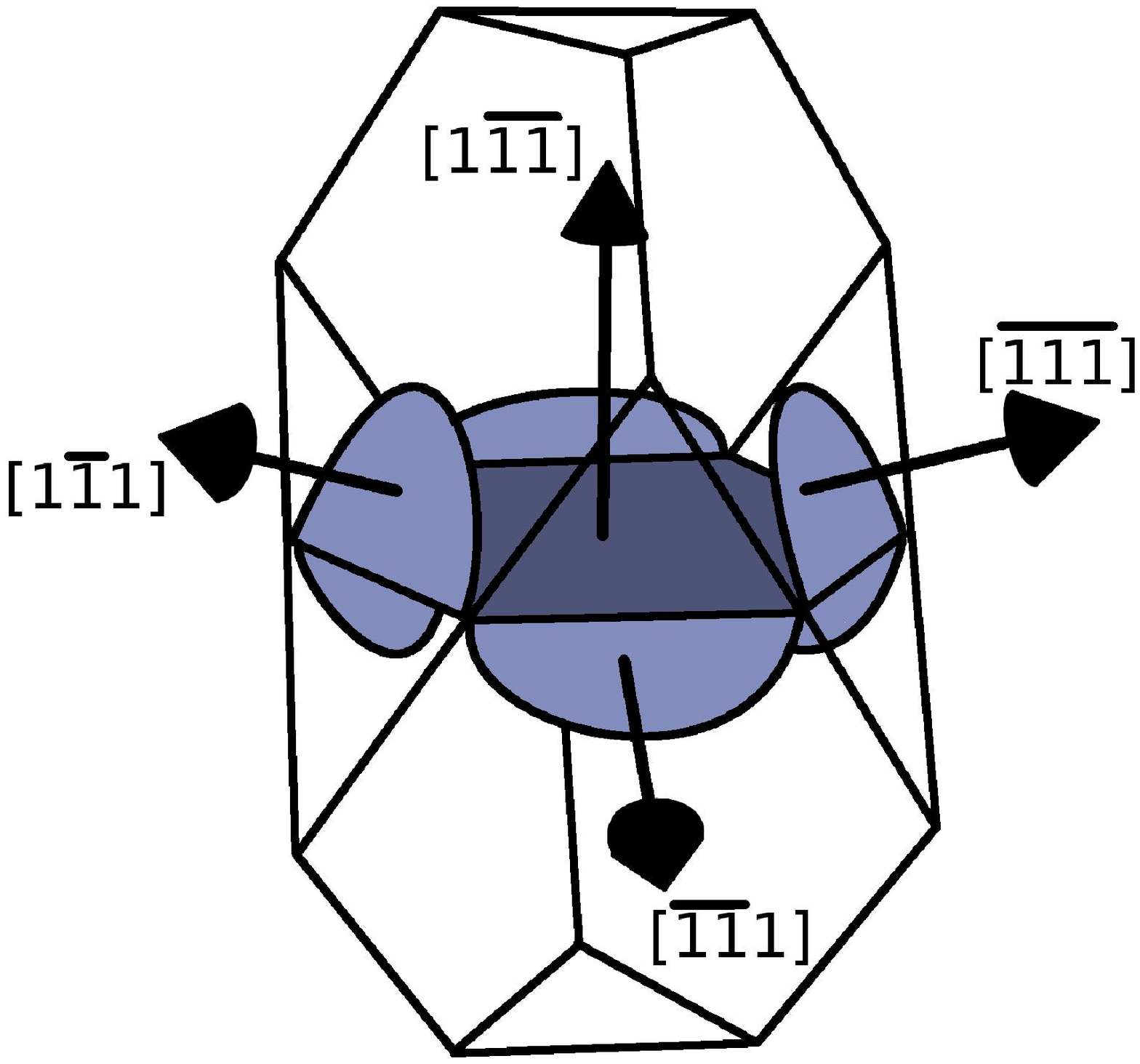}
\caption[The stacking fault surfaces formed on a dissociated Frank dislocation loop.]
{An illustration of the stacking fault surfaces formed on a dissociated Frank dislocation loop; the defect which provides a template for the formation of the tetrahedral 
bi-pyramidal precipitate assembly in the \alagA and \alagB alloys. 
The intersections of the \( \{111\} \) glide planes of the Shockley partial dislocations form a tetrahedral bi-pyramidal assembly whose edges are indicated by the thicker solid lines. 
The faulted regions generated by the central intrinsic fault and due to the passage of the 
Shockley partial dislocations are indicated by the darker and lighter shaded areas, respectively. 
 \label{fig-gamma-shockley}}
\end{center}
\end{figure}

Based on the present observations that the dislocation loop is dissociated and that the \gammaprime plates are formed both parallel to the loop and as well as on its edges it can be inferred that the \gammaprime phase nucleates on the stacking faults both in the plane of the Frank loop, 
and between the plane of the loop and the glissile Shockley partial dislocations. 
The formation  of the Shockley partial dislocations (described above) and their glide is repeated at each edge of the hexagonal Frank dislocation loop.
Adjacent Shockley partial dislocations lie on alternate faces of the habit plane of the dislocation loop, 
as illustrated in Figure~\ref{fig-gamma-shockley}, 
giving rise to the tetrahedral bipyramidal precipitate assemblies observed in this alloy. 
The closed assembly morphology adopted by the \gammaprime precipitates is therefore determined by the geometry of  the stacking fault  provided by the dissociated Frank dislocation loop.  

The separation, \(r\), of the Shockley and stair-rod partials can be estimated 
(neglecting anisotropy) using the equation\cite{clarebrough:1969a};
\[
r=Gb^2\frac{2+\nu}{36\pi\gamma_{sf}(1-\nu)} \label{eq-disoc-dist}
\]
where \(G\) is the shear modulus, \(b\) is the nearest-neighbor distance of the atoms, 
\(\gamma_{sf}\) is the stacking fault energy and \(\nu\) is Poisson's ratio.
While both silver and aluminium adopt face-centred cubic (fcc) structures, 
the stacking fault energy of silver is low (\(\mathrm{21.6\,mJ/m^2}\)
\cite{cahn_sfe})
 compared to aluminium (\(200\mathrm{\,mJ/m^2}\) \cite{gallagher:1970}).
Since the silver rapidly segregates to the GP zones the matrix has been shown to behave as a high stacking fault material with isolated low-stacking fault GP zones\cite{baur:1962}.
Recent theoretical work using Vienna \textit{ab initio} simulation package ({\scshape vasp})
 calculations have shown that silver segregation to stacking faults in aluminium is energetically favourable 
\cite{finkenstadt:2006}.
Earlier calculations indicated that increasing the silver content in aluminium alloys would  initially increase, 
then rapidly decrease the stacking fault energy\cite{schulthess:1998}. 
These calculations suggested that the stacking fault energy would fall below zero 
(favouring a hcp structure over fcc) for
 compositions in the range \(\sim\)45-90at\% Ag\cite{schulthess:1998}. 
This is significant since the composition of the \gammaprime phase falls within this range.
Based on these calculations, 
the presence of the stacking fault within the Frank loop would therefore lead to solute diffusion to the fault surface.  
This would also increase the separation between the Shockley and stair-rod partial dislocations, while simultaneously supplying solute to the growing precipitate. 
Even in the early stage of ageing, 
the region close to the precipitate assemblies were deficient in silver-rich GP zones compared to the surrounding matrix (see Figure~\ref{hrtem-1}), 
suggesting that solute from the GP zones has been incorporated into the precipitates.

The continued growth of the peripheral precipitates would lead to adjacent precipitates impinging on one another. 
This could lead to growth of one such plate and the expense of its neighbours.
In fact, with foils aged greater than 16 hours it was rare to find assemblies with all six peripheral precipitates present, with two plates (both larger than the central plate) on either face of the central plate being most the common arrangement, 
indicating precipitate coarsening within a given assembly. 

\paragraph{The effect of the bounding defect on  the lengthening of the \(\boldsymbol{\gammaprime}\) phase \label{ssec-bounding-defect}}

The central and peripheral \gammaprime precipitates were found to differ in their lengthening behaviour.
While the central plates grew rapidly to a diameter equivalent to the Frank dislocation loop, further lengthening was not observed.
The peripheral plates, which were initially much smaller than the central plates, 
were observed to lengthen and eventually had diameters greater than the central plates. 
The peripheral plates lengthened in only one direction.
This can be explained through the different partial dislocations bounding each precipitate.
As indicated in Figure~\ref{fig-frank-disoc}, 
the stacking fault on which the central precipitates of the assembly form is bounded by stair-rod dislocations (\burger{a}{6}{110}).
The peripheral precipitates are bounded by  stair-rod partial dislocations at their intersection with the central plate, 
and by the glissile Shockley partial dislocation (\burger{a}{6}{112}) on the opposite side. 
It should be noted (Figure~\ref{fig-frank-disoc-b}) that the stair-rod dislocation does not lie in the plane of either stacking fault, (the \plane[]{111} and \plane[]{11\overline{1}}planes in this figure) and is sessile.

The inability of the \gammaprime phase to grow beyond the stair-rod dislocation can be accounted for by the defect structure of the precipitate-matrix interface.
The lengthening mechanism of the \gammaprime phase has been proposed to involve the lateral motion of kinked ledges across the matrix-precipitate\cite{aikin:1987}, 
as is common for plate-shaped precipitates\cite{shiflet:1998}.
Each such ledge is associated with a transformation dislocation describing the displacement required to transform the matrix unit cell to that of the precipitate.
The structural component of the phase transformation can be described as the lateral glide of these ledges.
Each ledge of the \gammaprime precipitate has displacements equivalent to one or more Shockley partial dislocations, all of which lie in the habit plane.
The stair-rod dislocation, however, lies at the interface between the central and peripheral \gammaprime precipitates
and lies at an angle of 145\dg to the plane of the central precipitate. 
In this location it will displace the matrix and precipitate plane in the direction normal to the habit plane of the precipitate.
This dislocation is also sessile since the Burgers vector is not an atomic vector,  
despite lying on a glide plane. 
The present observations can tus be explained by this dislocation acting as a pinning site, 
inhibiting the glide of the transformation dislocation in the habit plane 
and thus the lengthening of the precipitate.

The high-resolution TEM images sometimes showed the presence of an additional 
peripheral precipitate, 
as 
at points \(B\) and \(C\) in Figure~\ref{hrtem-1b}.
The most probable explanation is that these smaller precipitates formed on jogs in the dislocation line.
Such jogs are thought to form during the condensation of additional vacancies onto dislocation loops\cite{silcock:gold}. 
Vacancy condensation on the loops would also be responsible for the  increase in the dislocation loop diameter in the \alagA alloy from \(\approx\)50\,nm in the as-quenched condition to \(\approx\)100\,nm after 0.083\,h ageing.  
The size of these precipitates is constrained by the jog and their growth is expected to be very limited.

\textbf{Comparison with previous results}

Given the amount of research conducted on the aluminium-silver alloy system it is perhaps surprising that the tetrahedral bi-pyramidal assemblies shown here have not been reported previously.
However, 
structured assemblies reported here were only observed in the absence of alternative nucleation sites and in particular those that acted as vacancy sinks.
It appears likely that in much of the earlier work, 
the presence of dislocations (often deliberately introduced) 
prevented the formation of such assemblies.
In the study performed by Frank \textit{et al.}\cite{frank:1961}, 
it is likely that such structures were formed,
but could not be effectively distinguished given the 
instrumentation available.

Grain boundaries regions were noticeably devoid of \gammaprime precipitates. 
Electron microprobe studies carried out on Al-Ag alloys  found that the width of the solute depleted region across a grain boundary was negligible compared to the 2\,\micron wide precipitate free zone, 
indicating that the precipitate free zone was a result of vacancy depletion, rather than a deficiency of solute \cite{clark:pfz}.  
This also explains the low density of \gammaprime precipitates reported in  Al-10.8wt\%Ag subjected to  equal channel angular pressing (ECAP)\cite{ohashi:2006}. 
The  grain size of the ECAP-treated samples  was of similar magnitude to the width of the precipitate free zones observed in the present work and consequently a considerable proportion of the vacancy population would have been lost from the sample, precluding the formation of the \gammaprime phase. 

The potential for forming 5-7 \gammaprime precipitates on each dislocation loop might appear to offer a potential means of increasing the number density, and thus enhance the precipitation strengthening, 
in this alloy.
However, the relatively low hardness values measured show that this is not the case. 
This is due to the fact that the precipitate are clustered together rather than uniformly distributed, 
an issue that has been examined and will be addressed in a separate work.

\section*{Conclusions}
\addcontentsline{toc}{section}{Conclusions}

The \gammaprime (\gammachem) phase was found to form on dissociated Frank vacancy condensation loops. 
When nucleated on this type of defect the precipitate forms hitherto unreported structured assemblies comprised of a central plate (formed on the surface of the Frank loop) and up to six peripheral precipitates growing on alternative \planes{111} planes from the faceted edges of the central plates.
This yields a three-dimensional assembly structure approximating a tetrahedral bipyramidal. 
HRTEM studies showed that the formation of the structures occurred very early during the ageing process.

The residual stair-rod dislocation bounding the Frank dislocation loop was a barrier to precipitate lengthening.
The diameter of the central plate did not increase, while the peripheral plates grew only away from the habit plane of the central plate. 

The heterogeneous nucleation site strongly influenced the microstructure,
controlling both the formation of the assembly structures and the subsequent lengthening of individual precipitates.

\textbf{Acknowledgments}
JMR gratefully acknowledges receipt of an Australian Research Council APA scholarship, and support through the ARC Centre of Excellence for Design in Light Metals.

\end{document}